\documentclass[aps,prl,twocolumn,showpacs,groupedaddress]{revtex4}  
\usepackage{graphicx}  
\usepackage{graphics}
\usepackage{dcolumn}   
\usepackage{bm}        
\usepackage{amssymb}   

\def\bra#1{\langle #1|}
\def\ket#1{| #1\rangle}
\def\HH{{\cal H}}
\def\LL{{\cal L}}
\def\NN{{\cal N}}

\def\PP{{\cal P}}
\def\braket#1#2{\langle #1|#2\rangle}
\def\moyenne#1{\langle #1\rangle}

\begin{document}



\title{Regulation by small RNAs via coupled degradation: mean-field and variational approaches}

\author{Thierry Platini}
\email{platini@vbi.vt.edu}
\affiliation{Virginia Bioinformatics Institute, Virginia Polytechnic Institute and State University,\\ Blacksburg, VA 24061, USA}

\author{Tao Jia}
\email{tjia@vt.edu}
\affiliation{Department of Physics, Virginia Polytechnic Institute and State University,\\ Blacksburg, VA 24061, USA}

\author{Rahul V. Kulkarni}
\email{kulkarni@vt.edu} 
\affiliation{Department of Physics, Virginia Polytechnic Institute and State University,\\ Blacksburg, VA 24061, USA}

\date{\today}

\begin{abstract}
Regulatory genes called small RNAs (sRNAs) are known to play critical
roles in cellular responses to changing environments. For several
sRNAs, regulation is effected by coupled stoichiometric
degradation with messenger RNAs (mRNAs).  The nonlinearity inherent in
this regulatory scheme indicates that exact analytical solutions for
the corresponding stochastic models are intractable.  Here, we present
a variational approach to analyze a well-studied stochastic model for
regulation by sRNAs via coupled degradation. The proposed approach is  
efficient and provides accurate estimates of mean mRNA levels as well as higher 
order terms. Results from the variational ansatz are in excellent agreement with data from stochastic simulations for a wide range of parameters, including regions of parameter space where mean-field approaches break down. The proposed approach can be applied to quantitatively model stochastic gene expression in complex regulatory networks.
\end{abstract}

\pacs{87.10.Mn, 02.50.r, 82.39.Rt, 87.17.Aa, 45.10.Db}
\maketitle 

A new paradigm for cellular regulation has emerged in recent years
with the discovery of novel non-coding genes called small RNAs
(sRNAs). In bacteria, sRNAs often function as global regulators that
mediate cellular adaptation to changing environments
\cite{gottesman05}. In higher organisms, the corresponding genes
(microRNAs) are known to play key roles in the regulation of critical
processes such as development, stem cell pluripotency and cancer
\cite{inui09,hornstein06}. It has been proposed that one of the key
functions of sRNAs in controlling cellular processes is to regulate
the variability (noise) in gene expression \cite{hornstein06}. Recent
experimental developments have led to approaches for quantifying such
variability using single-molecule measurements of mRNA
levels \cite{raj09}. These technological advances have now made
possible experimental studies that analyze the roles of sRNAs in noise
regulation during important processes such as development.
Correspondingly, there is a need for theoretical approaches that
complement such experimental efforts to enable a quantitative
understanding of different mechanisms of sRNA-based regulation.

While the molecular mechanisms of sRNA-mediated regulation continue to
be investigated, one established mechanism, representative of several
bacterial sRNAs, corresponds to binding with mRNAs followed by coupled
stoichiometric degradation \cite{masse03}.  An important challenge for
current research is to analyze how this regulatory mechanism impacts
the variability of gene expression across a population of cells.
Several recent theoretical studies \cite{levine08,levine07,mitarai07,mehta08,zhadanov09,shimoni07} have analyzed
models based on the corresponding reaction scheme (shown in
Fig. 1A).  The nonlinearity inherent in this reaction scheme implies
that exact analytical solutions for the corresponding stochastic model
are intractable; thus approximate analytical approaches are needed.
Previous theoretical studies have primarily focused on mean-field (MF)
approaches and on steady-state distributions using expansions around
MF solutions.  However, MF approaches are not accurate when we have a combination 
of nonlinear reaction rates (due
to interaction with small RNAs) and low mRNA/sRNA levels, which
points to the need for development of alternative analytical approaches.

In this paper, we analyze stochastic models of sRNA-based regulation
via coupled degradation (as shown in Fig. 1A). We first discuss the
MF approximation, which corresponds to neglecting mRNA-sRNA
correlations, and define dimensionless variables that are useful in
quantifying deviations between MF results and data from
stochastic simulations. To go beyond MF, we use a variational
approach which has been successfully applied to gene regulatory
networks in recent work \cite{sasai03,lan06b,Ohkubo07,Ohkubo08}.
Within this approach, we present a general ansatz for the steady-state
probability distribution which, at the simplest level, reduces to the
MF approximation. At the next level, the variational ansatz
gives results that are in excellent agreement with data from
simulations for the mean and variance of the regulated mRNA
distribution. The proposed method can be used for efficient and accurate quantitative 
analysis of sRNA-based regulation of gen expression.
\begin{figure}[tb]
\centering
\begin{center}
$\begin{array}{c@{\hspace{0.2cm}}c}
\multicolumn{1}{l}{\mbox{\bf (A)}} &
\multicolumn{1}{l}{\mbox{\bf (B)}} \\ [-0.0cm]
     \resizebox{30.0mm}{!}{
     \includegraphics*{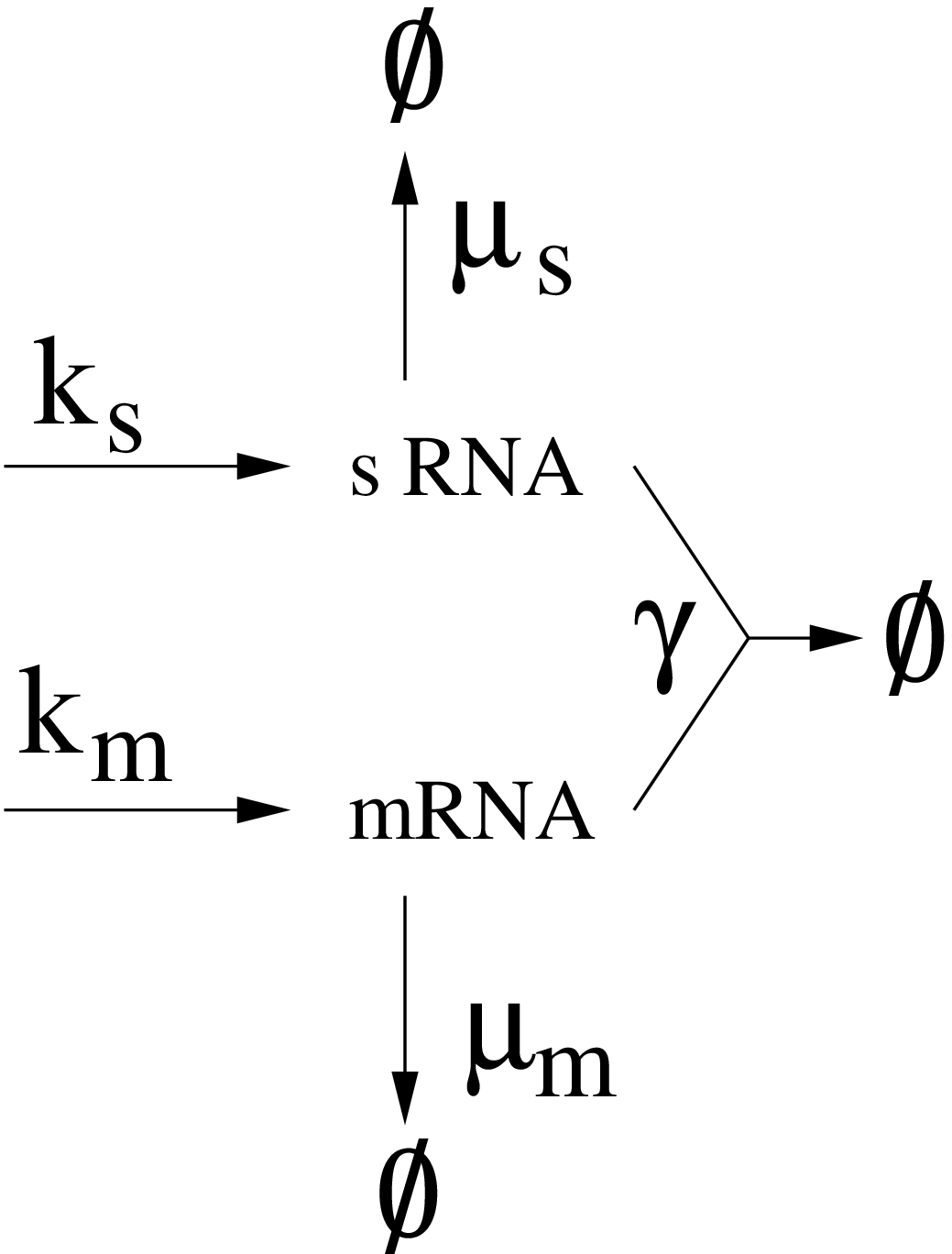}} &
     \resizebox{55mm}{!}{
     \includegraphics*{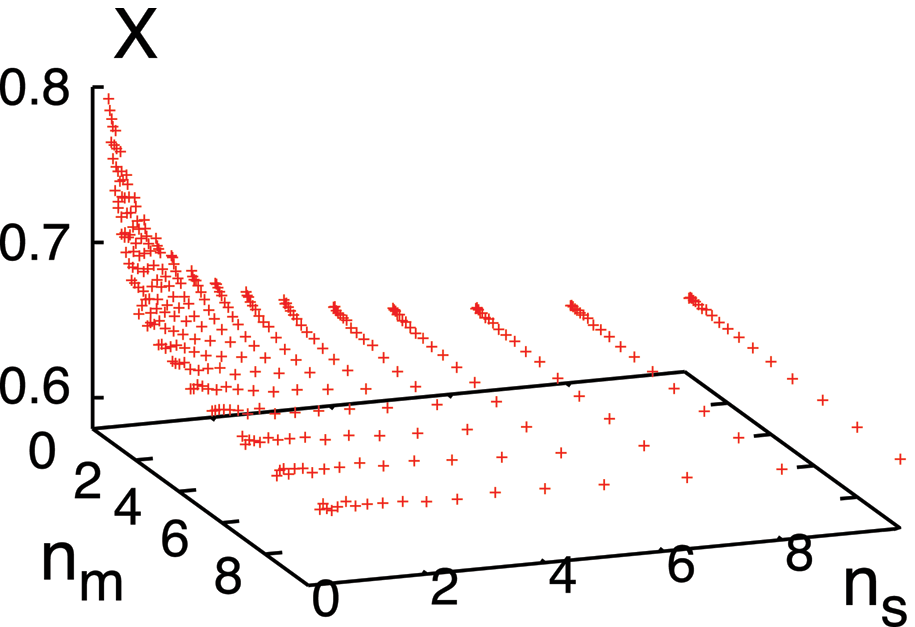}}\\
\end{array}$
 \caption{A) The kinetic scheme for regulation of mRNA by small RNAs
with coupled degradation rate $\gamma$. B) The ratio $X=\moyenne{m}/n_m$, obtained from 
simulation data, is
plotted as a function of $n_m$ and $n_s$. Parameters are chosen such that
$\epsilon_m=\epsilon_s=1$ and $\gamma=1$.  For $n_m,n_s\gg1$, $X$ converges towards the
MF preduction ($X\simeq0.618$).}
\end{center}
\label{Schema_and_X}
\end{figure}

We begin by considering the kinetic scheme presented in
Fig. 1A. The probability distribution of mRNA and
sRNA levels per cell, $P_{m,s}(t)$, obeys the master
equation:
\begin{eqnarray}\label{MasterEq1}
\partial_t P_{m,s}&=&k_mP_{m-1,s}+k_sP_{m,s-1}\\
&+&\mu_m(m+1)P_{m+1,s}+\mu_s(s+1)P_{m,s+1}\nonumber\\
&+&\gamma(m+1)(s+1)P_{m+1,s+1}\nonumber\\
&-&\left(k_m+k_s+\mu_mm+\mu_ss+\gamma ms\right)P_{m,s},\ \ \nonumber
\end{eqnarray}
where $k_j$, $\mu_j$ ($j=m,s$) and $\gamma$ are the parameters defined
in Fig. 1A. We will focus on the stationary distribution, denoted by $P^*_{m,s}$. It is convenient 
to define the following set of independent dimensionless parameters:
$\epsilon_m = k_s\gamma/\mu_m\mu_s$, $\epsilon_s = k_m\gamma/\mu_m\mu_s$ and $n_j=k_j/\mu_j$ ($j=m,s$).  
From the master equation
($\ref{MasterEq1}$), we can explicitly relate the average mRNA and sRNA levels to the correlation term $\moyenne{ms}$ \cite{elgart_bj,Levine_pnas} via:
\begin{eqnarray}\label{EQ_ms}
\frac{1}{\epsilon_m}\left(1 -\frac{\moyenne{m}}{n_m}\right) =\frac{1}{\epsilon_s}\left(1 -\frac{\moyenne{s}}{n_s}\right) = \frac{\moyenne{ms}}{n_m n_s},
\end{eqnarray}
where $\moyenne{.}$ denotes the stationary average. More generally, moments at one level are coupled to higher-order moments due to the nonlinear interaction term. This hierarchy makes the exact solution of the master equation intractable. Defining $X=\moyenne{m}/n_m$, $Y=\moyenne{s}/{n_s}$ and  
$C=\moyenne{ms}/\moyenne{m}\moyenne{s}$, equation (\ref{EQ_ms}) leads to
\begin{eqnarray}
\frac{1-X}{\epsilon_m} =\frac{1-Y}{\epsilon_s} = C \hphantom{.} XY.
\end{eqnarray} 

Traditionally, a first approximation, known as the
MF approximation, consists of neglecting correlations
through the substitution $\moyenne{ms}\rightarrow\moyenne{m}\moyenne{s}$. 
The MF assumption thus corresponds to $C=1$ and leads to
\begin{eqnarray}\label{MF-EQ}
\epsilon_m X Y + X -1 = 0, \ \ \ \epsilon_s X Y + Y -1 = 0.
\end{eqnarray} 
Comparing Eqns. (3) and (4), we see that the {\em exact} means (i.e. solutions of Eqn. (3)) are generated by the MF solutions considered with the rescaled interaction parameter $\gamma^{\prime} = C \gamma$.
Determination of $C$ can therefore provide accurate estimates of the mean mRNA and sRNA levels.
The ratio $C$ is also an indicator of the accuracy of MF: it is a good approximation when $C\simeq1$, whereas deviations from unity indicate that better approximations are needed. 

Furthermore, note that $X$ and $Y$ are, in general, functions of the four
parameters $\epsilon_m$, $\epsilon_s$, $n_m$ and $n_s$; however the
MF approximation (Eq. 4) predicts that both quantities depend only on
$\epsilon_m$ and $\epsilon_s$. It follows that MF theory breaks down
in regions of parameter space where $X$ and $Y$ depend on
the parameters $n_m$ and $n_s$ (for fixed $\epsilon_m$ and
$\epsilon_s$). These regions are indicated by significant deviations between the exact ratio $X$
($Y$) and the solution $\lambda_+$ ($\lambda_-$) of Eq. (\ref{MF-EQ}). 

We now anlayze deviations of the MF results from stochastic simulations data obtained using the Gillespie algorithm \cite{gillespie77}. The ratios $X$ and $C$ are plotted in figures 1B and 2A respectively.
These data are presented as a function of $n_m$ and $n_s$, keeping $\epsilon_m$ and $\epsilon_s$ constant.
The figures indicate that both quantities converge towards the MF predictions in the limit $n_s,n_m\gg1$ ($X\rightarrow 0.618$ and $C\rightarrow 1$). 
More significantly, the data shows that MF is not a good approximation for small $n_m$ and $n_s$. This is important to note since, in several cellular systems, mRNA abundances can be low (i.e. $n_m$ is small) \cite{tyson09}. This indicates that more accurate approximation are needed in such cases.

Furthermore, in the uncorrelated approximation, the
stationary probability distribution can be written as the product of
Poisson distributions $\Pi_{\lambda_+}(m)\times\Pi_{\lambda_-}(s)$,
where $\Pi_x(n)=e^{-x}x^n/n!$. Defining the marginal distributions
$P^*_m=\sum_sP^*_{m,s}$ and $P^*_s=\sum_mP^*_{m,s}$, the ratio $d_j
={\moyenne{j}}/({\moyenne{j^2}-\moyenne{j}^2})$ ($j=m,s$) measures
deviations between the marginals ($P^*_j$) and the simple Poisson
distribution. Again, deviations of $D=d_s\times d_m$ from unity
reveal that both marginal probability distributions cannot be
approximated by the Poisson distribution. In Fig. 2B, stochastic simulations data indicate that the coefficient $D$ deviates significantly from one for large $n_m$ and $n_s$. This observation implies that higher-order terms, such as
$\moyenne{m^2}$ and $\moyenne{s^2}$ cannot be obtained using the MF prediction
$\moyenne{j^2}-\moyenne{j}^2=\moyenne{j}$ ($j=m,s$), even in regions
of parameter space for which the mean values are given accurately by the
MF approximation. Interestingly, it is for small parameter values
$n_j$ ($j=m,s$), for which the MF approximation does not give accurate
mean values, that $D$ converges to one.  This observation is an
indication that the Poisson distribution is in some way embedded in
the structure of $P^*_{m,s}$.
\begin{figure}[tb]
\centering
\begin{center}
$\begin{array}{c@{\hspace{0.0cm}}c}
 \multicolumn{1}{l}{\mbox{\bf (A)}} \\  \multicolumn{1}{l}{\hspace{0.0cm}\resizebox{62.0mm}{!}{\includegraphics*{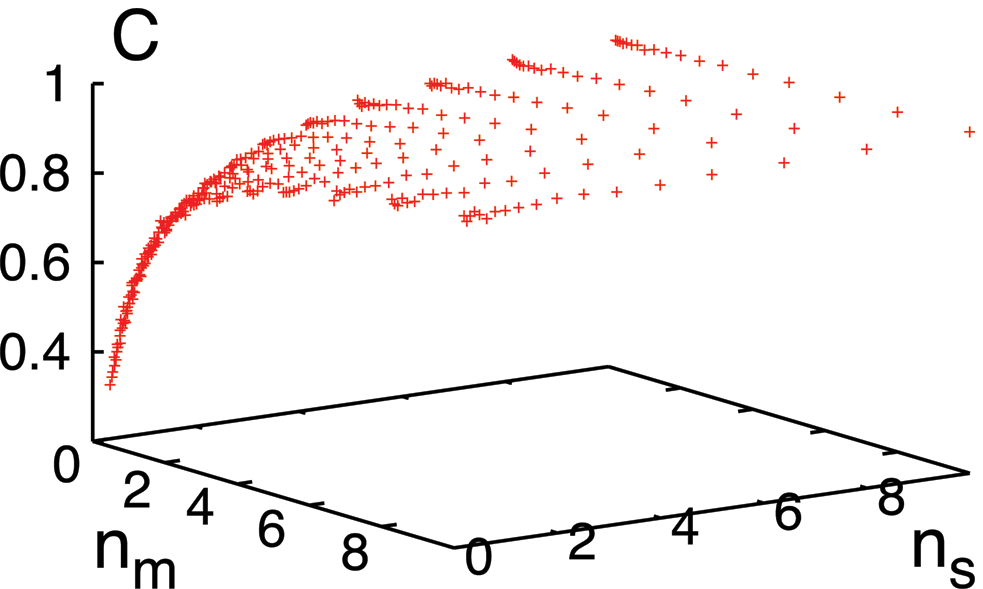}}}
\end{array}$ 
$\begin{array}{c@{\hspace{0.0cm}}c}
 \multicolumn{1}{l}{\mbox{\bf (B)}} \\  \multicolumn{1}{l}{\resizebox{64.0mm}{!}{\includegraphics*{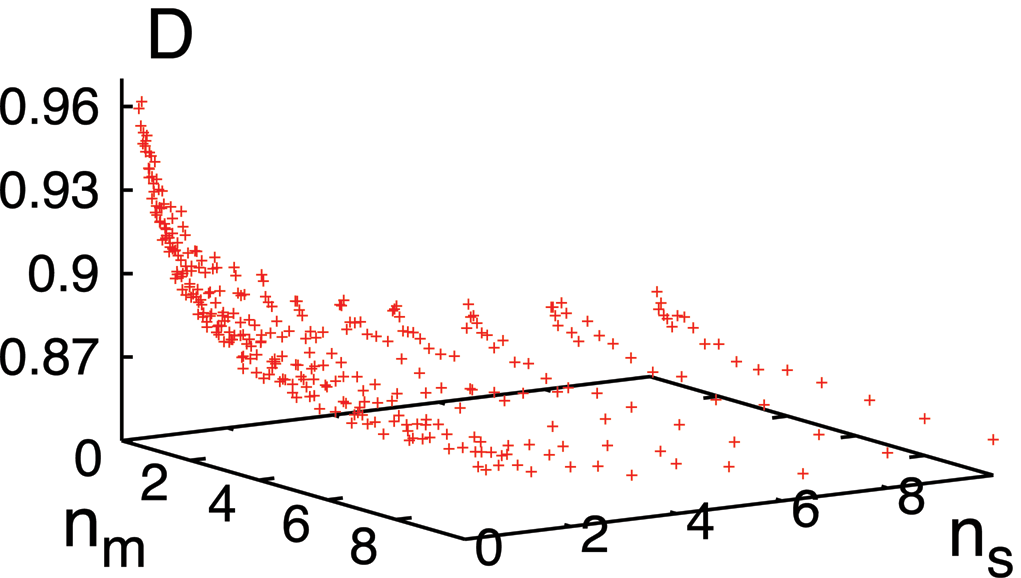}}}
\end{array}$
 \caption{Stationary value of $C=\moyenne{ms}/\moyenne{m}\moyenne{s}$ (A) and $D=d_m \times d_s$ (B), obtained from simulation data, plotted as a function of $n_m$ and $n_s$. We keep $\epsilon_m = \epsilon_s=1$ and $\gamma=1$.}
\end{center}
\label{FIG_Results}
\end{figure}

Based on the preceding analysis, it seems natural to
approximate $P^*_{m,s}$ as a superposition of Poisson distributions. 
This approximation can be implemented using 
the variational method introduced by Eyink
\cite{eyink}, combined with the quantum Hamiltonian formalism of the
master equation \cite{sasai03,lan06b}. Following the mapping outlined by 
Doi \cite{doi76}, we define the operators $a^\dag$
and $a$ (respectively $ b^\dag$ and $b$) associated with the creation
and annihilation of mRNA (sRNA). The master equation (\ref{MasterEq1})
takes the compact form $\partial_t \ket{\psi(t)}=-\LL\ket{\psi(t)}$ with
\begin{eqnarray}
\LL&=&k_m+k_s+\mu_m a^\dag a+\mu_s b^\dag b+\gamma a^\dag ab^\dag b\nonumber\\
&-&\left(k_ma^\dag+k_sb^\dag+\mu_ma+\mu_sb+\gamma ab\right),
\end{eqnarray}
where $[a,a^\dag]=[b,b^\dag]=1$ and $a\ket{0,s}=0$,
$b\ket{m,0}=0$. Focusing on the stationary state, we denote by $\bra{\psi_L}$ and $\ket{\psi_R}$ the left and right
eigenstates with vanishing eigenvalue. They obey $\braket{\psi_L}{n,m}=\braket{\psi_L}{\psi_R}=1$. The mapping to the original problem is given by $P^*_{m,s}=\braket{m,s}{\psi_R}/m!s!$.

To initiate the variational ansatz, we define the left and
right trial vectors ($\bra{\phi_L(\Lambda_L)}$ and
$\ket{\phi_R(\Lambda_R)}$), constructed using a set of
independent parameters, $\Lambda_L$ and $\Lambda_R$. Defining the
functional $\HH(\Lambda_L, \Lambda_R)=\bra{\phi_L}\LL\ket{\phi_R}$,
the eigenstates are determined using the variational principle
$\delta\HH=0$. A detailed explanation of the variational scheme is provided in \cite{eyink}.

We now generalize the uncorrelated approximation to propose a
specific ansatz for the trial vectors as the superposition of Poisson
distributions. A similar ansatz has also been proposed in a recent
study of reaction systems including different chemical species
\cite{Ohkubo08}. We define
\begin{eqnarray}
\bra{\phi_L(\Lambda_L)}&=&\bra{0,0}e^{a+b}\prod_{i,j=0}^{d}e^{\theta_{i,j}a^ib^j},\\
\label{RightEigenV}
\ket{\phi_R(\Lambda_R)}&=&\sum_{i,j=1}^{d}\Theta_{i,j}e^{\alpha_i(a^\dag-1)}e^{\beta_j(b^\dag-1)}\ket{0,0},
\end{eqnarray}
with $\Lambda_R=\{\alpha_p,\beta_q,\Theta_{p,q}\}$ and
$\Lambda_L=\{\theta_{p,q}\}$ ($\theta_{d,d}=0$). In each vector, the
total number of parameters $\NN$ is given by
$\NN=d(d+2)$. The parameters of $\bra{\phi_L}$ are
imposed by the condition $\braket{\phi_L}{m,n}=1$ which leads to $\theta_{p,q}=0, \forall p,q$.
It follows that the set $\Lambda_R$ is solution of $\bra{\delta\phi_L}\LL\ket{\phi_R}|_{\Lambda_L=\{0\}}=0$. Our calculation leads to the system of equations:
\begin{eqnarray}\label{SetofEq}
&\sum_{p,q=1}^d \Theta_{p,q}\alpha_p^i\beta_q^j\times\left[
\epsilon_s \epsilon_m(ij+i\beta_q+j\alpha_p)
\right.
\\ &\left.+ 
in_s \epsilon_s(1-n_m/\alpha_p)+jn_m \epsilon_m(1-n_s/\beta_q)
\right]=0,\nonumber
\end{eqnarray}
generated for $i,j=0,1,2,..., d$ with the pair $(i=d,j=d)$ excluded.
The first equation (for $i=j=0$), corresponds to the probabilistic
interpretation: $\braket{\phi_L}{\phi_R}=1$ and leads to the normalization constraint $\sum_{p,q}\Theta_{p,q}=1$.  From equation ($\ref{SetofEq}$) one can then generates the $\NN$ independent conditions required to determine the right eigenvector parameters. It follows that an approximation of the stationary distribution is given by $\PP_{m,s}^*=\braket{m,s}{\phi_R(\Lambda_R^*)}/m!s!$, where $\Lambda_R^*=\{\alpha_p^*,\beta_q^*,\Theta_{p,q}^*\}$ is solution of ($\ref{SetofEq}$).
The latter distribution can be explicitly written as a superposition of Poisson distributions: $\PP^*_{m,s}=\sum_{p,q}\Theta^*_{p,q}\Pi_{\alpha^*_p}(m)\Pi_{\beta^*_q}(s)$. We note that the MF results are recovered by considering the ansatz with $d=1$. In this case, $\PP^*_{m,s}$ is simply a product of two Poisson distributions with means $\alpha$ and $\beta$ respectively. The variational equations give $n_s(n_m-\alpha)-\epsilon_m\alpha\beta=0$ and $n_m(n_s-\beta)-\epsilon_m\alpha\beta=0$, leading to $X=\lambda_+$ ($Y=\lambda_-$) and $C=D=1$.
\begin{figure}[tb]
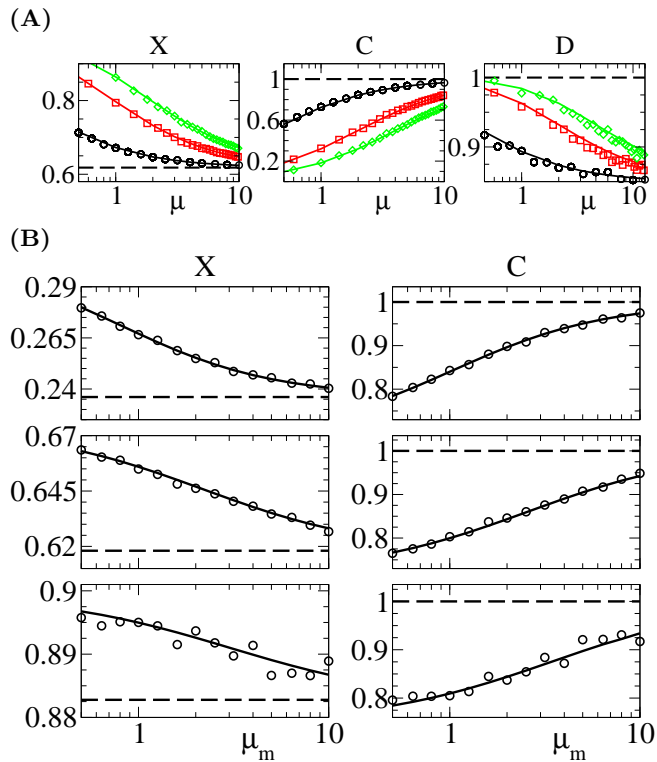

\centering
\begin{center}
$\begin{array}{c@{\hspace{0.0cm}}c}
 \multicolumn{1}{l}{\mbox{\bf (A)}} \\  \multicolumn{1}{l}{\hspace{0.42cm}\resizebox{81.0mm}{!}{\includegraphics*{FIG_XCD_Sym_Final.eps}}}
\end{array}$ 
$\begin{array}{c@{\hspace{0.0cm}}c}
 \multicolumn{1}{l}{\mbox{\bf (B)}} \\  \multicolumn{1}{l}{\resizebox{86.0mm}{!}{\includegraphics*{FIG_XD_Non_Sym_Final.eps}}}
\end{array}$
 \caption{Comparisons of simulation data (symbols), ansatz predictions (lines) and MF results (dashed line). (A) The quantities $X=\moyenne{m}/n_m$, $C=\moyenne{ms}/\moyenne{m}\moyenne{s}$ and $D=d_s\times d_m$ are plotted as a function of $\mu=\mu_m=\mu_s$ on a logarithmic scale, for $\gamma=1$ (circles), $\gamma=5$ (squares), and $\gamma=10$ (diamonds). We keep $\epsilon_m = \epsilon_s =1$ with $k_m=k_s=k$. (B) The quantities $X$ (left) and $C$ (right) are plotted as a function of $\mu_m$ on a logarithmic scale, for $\epsilon_m =4$ (top), $\epsilon_m =1$ (middle) and $\epsilon_m =1/4$ (bottom). We keep $\mu_s=2$, $\gamma=1$ and $\epsilon_s =1$.}
\end{center}
\label{FIG_Results}
\end{figure}

Going one step beyond the MF approximation, we
consider the ansatz ($\ref{RightEigenV}$) with $d=2$.
We first consider the symmetric case $k_m=k_s=k$ and $\mu_m=\mu_s= \mu $. This choice imposes
$\alpha_j=\beta_j$ ($j=1,2$) and $\Theta_{1,2}=\Theta_{2,1}$. The set $\Lambda_R^*$,
solution of the equations generated by (\ref{SetofEq}), is obtained
numerically using standard routines. From a practical point of view,
the numerical calculation is significantly faster
than stochastic simulations, especially if we need to explore
large regions of parameter space.

Figure $3A$ presents a comparison of our results with data from stochastic simulations.  Keeping the ratios
$\epsilon_m$ and $\epsilon_s$ constant, the quantities $X$, $C$ and
$D$ are plotted as a function of $\mu$ for $\gamma=1$, $5$ and
$\gamma=10$. Clearly, deviations from MF results appear more pronounced as $\gamma$ increases. However, for a range of parameter values $\mu$ and even for large mRNA-sRNA coupling, the variational scheme gives accurate values
of the mean mRNA level per cell ($\moyenne{m}=X\times
n_m$). Additionally, we checked that the predictions for $\moyenne{s}$
also presents an excellent agreement with simulation data. Importantly, the agreement of our predictions with simulation data,
for the quantities $C$ and $D$, shows that the variational method also
gives accurate values of higher order terms such as the correlation
$\moyenne{ms}$ $(=C\times\moyenne{m}\moyenne{s})$ and variance
$\moyenne{j^2}-\moyenne{j}^2$ $(=\moyenne{j}/d_j)$.

To compare our results in the non-symmetric case, we
consider variations in $\mu_m$, keeping $\mu_s=2$ and $\gamma=1$
fixed. The set of parameters is once again computed numerically,
solving $8$ coupled equations generated from equation (\ref{SetofEq}). 
The ratio $\epsilon_s$ is kept equal to unity while
$\epsilon_m =4$, $1$ and $1/4$. As shown in Fig. $3B$, the ansatz predictions are, once again, in excellent agreement with simulation data.

In conclusion, we have presented a variational approach for analyzing a
coupled degradation mechanism of sRNA-based regulation. The latter method generates a set of algebraic equations that can be solved numerically. At the simplest level, the approach reduced to the MF approximation
which is shown to be inaccurate for low abundances of the interacting
components. The approach proposed allows for systematic improvements
over MF and, at the next level, gives excellent agreement with
simulation data for the mean and variance of steady-state mRNA/sRNA
distributions. The results derived will aid approaches for inference
of model parameters from experimental measurements of mean and
variance. More generally, the proposed approach can be extended to
treat other biological networks with nonlinear interactions for which
analytical solutions of the corresponding stochatsic models are
intractable. In such cases, the proposed procedure of constructing the variational ansatz  (i.e. superposition of MF probability distributions) can lead to accurate estimates of the mean and variance for quantities of interest.
It is hoped that future work coupling such approaches
with experiments will lead to quantitative understanding of gene expression in complex networks.

\noindent We would like to thank the Stat. Mech. and NDSSL groups at Virginia Tech, especially professor S. Eubank. This research is funded by the US National Science Foundation through PHY-0957430, DMR-0705152 and the NIH MIDAS project 2U01GM070694-7.

\bibliographystyle{apsrev}
\bibliography{stochastic_modeling}

\end{document}